\documentclass[aps,prl,twocolumn]{revtex4}
\usepackage{amssymb}
\usepackage{amsmath}
\usepackage{graphicx}
\usepackage{epsfig,amsmath}

\begin{document}

\title{Diffusion of Inhomogeneous Vortex Tangle and Decay of Superfluid
Turbulence }
\author{Sergey K. Nemirovskii}
\affiliation{Institute of Thermophysics, Lavrentyev ave., 1, 630090, Novosibirsk, Russia\\
Novosibirsk State University, Pirogova 2, 630090, Novosibirsk, Russia}

\begin{abstract}
The theory describing the evolution of inhomogeneous vortex tangle at zero
temperature is developed on the bases of kinetics of merging and splitting
vortex loops. Vortex loops composing the vortex tangle can move as a whole
with some drift velocity depending on their structure and their length. The
flux of length, energy, momentum etc. executed by the moving vortex loops
takes a place. Situation here is exactly the same as in usual classical
kinetic theory with the difference that the "carriers" of various physical
quantities are not the point particles, but extended objects (vortex loops),
which possess an infinite number of degrees of freedom with very involved
dynamics. We offer to fulfill investigation basing on supposition that
vortex loops have a Brownian structure with the only degree of freedom,
namely, lengths of loops $l$. This conception allows us to study dynamics of
the vortex tangle on the basis of the kinetic equation for the distribution
function $n(l,t)$ of the density of a loop in the space of their lengths.
Imposing the coordinate dependence on the distribution function $n(l,\mathbf{%
r},t)$ and modifying the "kinetic" equation with regard to inhomogeneous
situation, we are able to investigate various problem on the transport
processes in superfluid turbulence. In this paper we derive relation for the
flux of the vortex line density $\mathcal{L}(x,t)$. The correspoding
evolution of quantity $\mathcal{L}(x,t)$ obeys the diffusion type equation
as it can be expected from dimensional analysis. The according diffusion
coefficient is evaluated from calculation of the (size dependent) free path
of the vortex loops. We use this equation to describe the decay of the
vortex tangle at very low temperature. We compare that solution with recent
experiments on decay of the superfluid turbulence.
\end{abstract}

\maketitle

\section{\protect\bigskip Introduction and Scientific Background}

Idea that inhomogeneous vortex tangle evolves in a diffusive-like manner
appeared quite long ago. Thus, the authors of paper \cite{Beelen88}, who
observed the regions of \ high vortex line densities $\mathcal{L}(\mathbf{r,}%
t)$ --- \textquotedblleft plugs\textquotedblright\ in the channel with the
counterflowing He II proposed that this phenomenon appeared due to diffusion
of quantity $\mathcal{L}(\mathbf{r,}t)$. An attempt to describe
theoretically these processes was made by \cite{Beelen88}, and \cite%
{Geurst89}, who proposed to introduce the term proportional $\nabla ^{2}%
\mathcal{L}(\mathbf{r,}t)$ into the Vinen equation. Authors were not able to
restore the value of the diffusion coefficient from experiment (these
results are reviewed and discussed in \cite{NF1995}). In series of works
(see \cite{JM2007} and references therein) both the diffusion process and
the diffusion coefficient we obtained from the so called Non Equilibrium
Thermodynamics principles developed by the authors earlier. In paper \cite%
{Tsub-diff} \ the spatial diffusion of an inhomogeneous vortex tangle\ had
been studied numerically (see Fig. 1). \ Analyzing their results the authors
determined the diffusion constant to be of order of $0.1\kappa $ ($\kappa $
is the quantum of circulation). Dynamics of the inhomogeneous vortex tangle
had been studied numerically also in paper \cite{Barenghi-evap}, however,
since the authors studied the dilute tangle they observed the "ballistic"
regime rather then pure diffusion.
Especial interest to the diffusion processes arises in context of the
problem on decay of \ the vortex tangle at zero (extremely low) temperature.
Indeed, the most apparent mechanism of dissipation in quantum fluids - the
mutual friction between the vortices and the normal component -disappears.
At the same time in a series of experimental works \cite{mac-clintock},\cite%
{Pickett-2006},\cite{Golov07} the decay of quantum turbulence is \ observed,
and the question arises what is the mechanism for dissipation at zero
temperature. Various approaches and ideas such as a cascade like break down
of the loops, Kelvin waves cascade, acoustic radiation, reconnection loss
etc. have been discussed in details in recent review \cite{barenghi-decay}.
\ The mechanism of the vortex tangle speading with the subsequent
degeneration is usually ignored. In fact, contibution of diffusion had been
discussed in the cited experimental work, but grounding on the value $%
0.1\kappa $ for the diffusion constant, \ the authors concluded that this
small diffusion coefficient did not lead to correct time of decay. In the
present paper we develop the theory describing the evolution of an
inhomogeneous vortex tangle on the bases of kinetics of the merging and
breaking down vortex loops. \ We showed that evolution of a weakly
inhomogeneous vortex tangle obeys the diffusion equation with the
coefficient equal to adout $2.2\kappa $, which exceeds approximately
twentyfold as large the value obtained in \cite{Tsub-diff}. We present
arguments that the diffusion constant would be significantly underestimated
in \cite{Tsub-diff} due to especial procedure used by the authors. We used
the diffusion equation to describe the decay of the vortex tangle at very
low temperature. Comparison with the recent experiments on decay of the
superfluid turbulence \cite{mac-clintock},\cite{Pickett-2006},\cite{Golov07}
is made.

\section{The evolution of inhomogeneous vortex tangle}

\subsection{Free path}

Vortex loops composing the vortex tangle can move as a whole with some drift
velocity $V_{l}$ depending on their structure and their length. The flux of
the line length, energy, momentum etc., executed by the moving vortex loops
takes place. In the case of inhomogeneous vortex tangle the net flux due to
gradient of concentration of the vortex line density appears. Situation here
is exactly the same as in classical kinetic theory with the difference that
the "carriers" are not the point particles but the extended objects (vortex
loops), which possess an infinite number of degrees of freedom with very
involved dynamics. In addition, while collision (or self-intersection) of
elements of filaments the so called reconnection of the lines occurs, and
loops either merge or split, losing their individuality \ and turning into
other loops. The full statement of this problem requires some analog of the
secondary quantization method for extended objects, or the string field
theory, the problem of incredible complexity. Clearly, this problem can be
hardly resolved in the nearest future. Some approach crucially reducing the
number of degrees of freedom is required.

\begin{figure}[tbp]
\includegraphics[width=7cm]{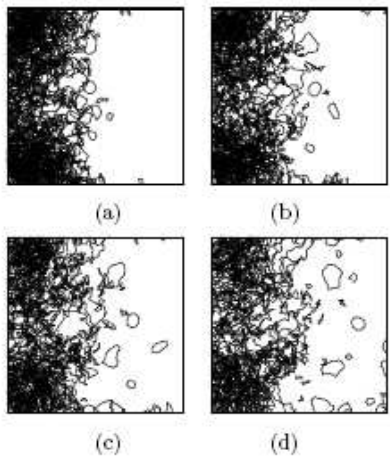}
\caption{Diffusion of a vortex tangle at t=0 sec(a), t=10.0 sec(b), t=20.0
sec(c) and t=30.0 sec(d),(Tsubota et, al.
\protect\cite{Tsub-diff}.}
\label{}
\end{figure}

We offer to fulfill investigation basing on supposition that vortex loops
have the Brownian structure (see \cite{nemirPRB1998}).{\Huge \ }In
accordance with this approach the average loop can be imagined as consisting
of many arches with the mean radius of curvature equal $\xi _{0}$ randomly
(but smoothly) connected to each other. Quantity $\xi _{0}$ is important
parameter of the approach. It plays a role of the "elementary step" in the
theory of polymer. It is low cut-off of the approach developed, the theory
does not describe scales smaller then $\xi _{0}$. Statistics of Brownian
loops is described by the generalized Wiener distribution with the only
parameter $\xi _{0}$ for all loops. Thus we consider the vortex tangle as a
collection of vortex loops of various lengths $l$, which is only degree of
freedom. Let us introduce distribution function $n(l,t)$ of the density of a
loop in the \textquotedblright space\textquotedblright\ of their lengths. It
is defined as the number of loops (per unit volume) with lengths lying
between $l$ and $l+dl$. Knowing quantity $n(l,t)$ and statistics of each
personal loop we are able to evaluate various properties of real vortex
tangle. The distribution function $n(l,t)$ obeys the Boltzmann type
"kinetic" equation. Study of exact solution to this "kinetic" equation
allowed us to develop \ a theory of superfluid turbulence, which
quantitatively describes the main features of this phenomenon \cite%
{NemirPRL2006},\cite{Nemir2008}.

This approach turns out to be useful for the study of inhomogeneous vortex
tangle. In this case we have to impose the coordinate dependence on the
distribution function and on parameter $\xi _{0}$, that is to put $n(l,%
\mathbf{r},t)$, $\xi _{0}(\mathbf{r},t)$, and to modify the "kinetic"
equation with regard to inhomogeneous situation. In fact, in this work we
restrict ourselves to a bit more modest problem, namely we study the
question of the spatial and temporal evolution of the vortex line density $%
\mathcal{L(}\mathbf{r,}t)$. The corresponding theory can be developed in
spirit of the classical \ kinetic theory with the difference \ that the
transport processes are executed with the extended objects - vortex loops.
Accordingly the key questions is to evaluate the drift velocity $V_{l}$ and
the free path for the loop of size $l$. The drift velocity $V_{l}$ is
defined via an averaged quadratic velocity of the line elements (simple
average velocity vanishes due to symmetry)
\begin{equation}
V_{l}=\sqrt{\frac{1}{l}\int \left\langle \mathbf{\dot{s}(\xi )}%
^{2}\right\rangle d\xi }.  \label{V_drift}
\end{equation}%
The averaging $\left\langle {}\right\rangle $ should be done with use of the
Gaussian model briefly described above \cite{nemirPRB1998}. The result of
according calculations is that $V_{l}=\beta /\sqrt{l\xi _{0}}$. Quantity $%
\beta $ is $C_{v}(\kappa /4\pi )\ln (l/a_{0})$, where $a_{0}$ is the core
radius and $C_{v1}$ is a constant about unity. Velocity $V_{l}$ can be also
estimated from the following qualitative consideration. Consequently
considering the average loop as consisting of $n=l/\xi _{0}$ arches with the
mean radius of curvature equal to $\xi _{0}$ randomly (but smoothly)
connected to each other, we take its velocity as the resulting velocity of
all arches composing the loop. Since the arches randomly connected to each
other, and have the\textit{\ }velocity as for rings\textit{, }$%
V_{arch}=\beta /\xi _{0}$ (directed along the normal), the resulting
averaged velocity is the "random walking" average, $V_{l}=\frac{1}{n}\sqrt{n}%
V_{arch}=\beta /\sqrt{l\xi _{0}}$.

The drift motion is realized until the loops collide with other loops with
subsequent formation of larger loops. Number of collisions $P_{Col}(dt)$ per
small interval $dt$ can be estimated from the "kinetic equation" for the
distribution function $n(l)$ of density of loops in space of their lengths $%
l $ (see \cite{NemirPRL2006},\cite{Nemir2008}). The rate of change of
density $n(l)$ due to collisions is%
\begin{equation}
\frac{\partial n(l,t)}{\partial t}=-2\int \int A(l_{1},l,l_{2})\delta
(l_{2}-l_{1}-l)n(l)n(l_{1})dl_{1}dl_{2}.\;  \label{KE_part}
\end{equation}%
We omit the processes of the loops breakdown due to the self-intersection.
The reason is that the migration of loops is performed mainly by small
loops, the large ones undergo reconnections without any essential drift. The
scattering cross-section $A(l_{1},l,l_{2})$ describes the rate (number of
events per unit volume and unit time) of collision of two loops with lengths
$l$ and $l_{2}$ and forming the loop of length $\ l_{1}+l=l_{2}$. It is
evaluated in papers \cite{nemir2006LTP}\ $\
A(l_{1},l,l_{2})=b_{m}V_{l}l_{1}l.$ Here $b_{m}$ is the numerical factor
approximately equal to $\ b_{m}\approx 0.2$ and $V_{l}$ $=\beta /\xi _{0}$
is the velocity of approaching of vortex line elements (similar result for
quantity $b_{m}\sim 0.1\div 0.2$ was obtained qualitatively in the context
of cosmic strings \cite{Copeland98}). Then probability $P_{Col}(dt)$ for
loop to collide with other loops (and reconnect) in small interval $\ dt$ is:

\begin{equation}
P_{Col}(dt)=\Lambda dt,  \label{P_collision}
\end{equation}%
where quantity $\Lambda $ is evaluated with the use of relation (\ref%
{KE_part})
\begin{equation}
\Lambda =2\int \int A(l_{1},l,l_{2})\delta
(l_{2}-l_{1}-l)n(l_{1})dl_{1}dl_{2}.\;  \label{Lambda}
\end{equation}%
We calculate the collision probability $\Lambda $ with the use of \ the
distribution function $n(l)$, obtained early $n(l)=Cl^{-5/2}$ (see \cite%
{NemirPRL2006},\cite{Nemir2008}). Here coefficient $C$ can be ascertained
from the normalization condition that the vortex line density $\mathcal{L}$
is just $\mathcal{L}=\int n(l)ldl$ (we used while integration the fact that
quantity $\xi _{0}$ serves as the low cut-off ). Simple calculation leads to
relation:
\begin{equation}
\Lambda (l)=2\beta b_{m}\mathcal{L}\sqrt{\frac{l}{\xi _{0}}.}
\end{equation}

In the usual way we conclude that probability $P(x)$ for the loop of length $%
l$ to fly distance $x$ without collision is%
\begin{equation}
P(x)=2l\mathcal{L}\exp (-2lb_{m}\mathcal{L}x).  \label{probability(x)}
\end{equation}
It can be seen from relation (\ref{probability(x)}) that the free path for
loop of length $l$ is $1/2lb_{m}\mathcal{L}$.

\subsection{The flux of length and the diffusion equation}

Knowing the $l$ and $\xi _{0}$-dependent averaged velocity of loops and the
probability $P(x)$ we can evaluate the spacial flux of the vortex line
density $\mathcal{L}$ executed by the loops. The procedure is very close to
the one made in the classical kinetic theory with the difference that the
carriers have different sizes, this requires additional integration over the
loop lengths.

\begin{figure}[tbp]
\includegraphics[width=5cm]{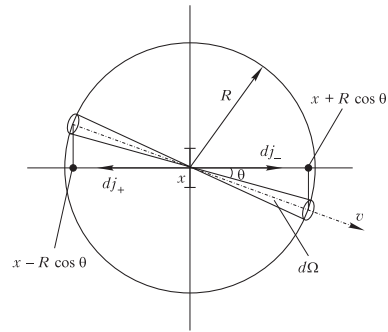}
\caption{The net flux of
length through the small area element placed in $x=0$ and orientated
perpendicularly to axis $x$}
\label{}
\end{figure}

Let us consider the small area element placed at some point $x$ and
orientated perpendicularly to axis $x$ (See Fig.2). The $x$-component flux
of the line length executed by loops of sizes $l$ placed in $\theta ,\varphi
$ direction (from the left and right sides, correspondingly) and remote from
the area element at distance $R$ can be written as $d\mathbf{j}_{\mp
}(\theta ,\varphi ,l,R)=l\ n(\theta ,\varphi ,l,R)V_{l}\cos \theta $ $P(R)$.
The net is%
\begin{equation}
\mathbf{J}_{x}=\int (d\mathbf{j}_{+}(\theta ,\varphi ,l,R)-d\mathbf{j}%
_{-}(\theta ,\varphi ,l,R))\frac{\sin \theta d\theta d\varphi }{4\pi }dldR.
\label{net flux}
\end{equation}%
Taking that spacial density of the loops is the function of coordinate $x$
(i.e. $n(l,x)$) we obtain after simple integration
\begin{equation}
\mathbf{J}_{x}=-\frac{1}{24}\frac{\beta }{\mathcal{L}b_{m}}\left[ \frac{1}{%
2\xi _{0}^{3}}\mathcal{L}\frac{\partial }{\partial x}\xi _{0}+\frac{1}{\xi
_{0}^{2}}\frac{\partial }{\partial x}\mathcal{L}\right] .  \label{flux final}
\end{equation}

There are possible the following three variants: i). The mean radius of
curvature $\xi _{0}$ is the fixed quantity (this case can be realized while
injecting the loops (rings) of some fixed size into volume.). ii).Quantity $%
\xi _{0}(x)\ $\ is independent variable, which can appear, when the own
vortex line (kelvin wave propagation) dynamics prevails collisions of loop,
determining thereby the structure of the vortex tangle. In this case the
mean radius of curvature $\xi _{0}$ should be connected to spectrum of
Kelvin waves\cite{Svistunov95} iii). Mean curvature $\xi _{0}(x)$ and vortex
line dinsity $\mathcal{L}(x)$ are not independent, the relation between
these two quantaties appears due to the balancing of fusion and splitting of
the loops and can be determined from the "kinetic" equation \cite%
{NemirPRL2006}, it is $\xi _{0}\approx \allowbreak 0.27\mathcal{L}^{-1/2}$.
The fact that the interline space is of the order of the mean radius of
curvature had been firstly discovered numerically in \cite{Schwarz88}. Using
this connection we come to conclusion that $\mathbf{J}_{x}=-D_{v}\partial
\mathcal{L}/\partial x$ and correspondingly, the spatial-temporal evolution
of quantity $\mathcal{L}$ obeys the diffusion type equation%
\begin{equation}
\frac{\partial \mathcal{L}}{\partial t}=D_{v}\nabla ^{2}\mathcal{L},
\label{diffusion equation}
\end{equation}

where the diffusion coefficient $D_{v}$ is equal $C_{d}\kappa $. Our
approach is fairly crude to claim for good quantitative description.
However, if to adopt the data grounded on exact solution \ to the Boltzmann
type "kinetic" equation (\cite{Nemir2008}) we conclude that $C_{d}\approx
2.2 $. As it was discussed above, the result that evolution of VLD obeys the
diffusion type equation with the diffusion constant of order of $\kappa $ is
expected. The actual interest is connected to prefactor $C_{d}$. Early, the
only quantitative result was obtained numerically by Tsubota et al. , and it
was $C_{d}\approx 0.1$, in $22$ times smaller. We will discuss the probable
reason for this large discrepancy later.

\section{Boundary conditions}

To construct the boundary conditions we have to state the problem
specifically.

{\large 1. Smearing of the tangle. } Let us consider, first, the case when
the vortex tangle is placed in some restricted domain of superfluid helium.
Let us consider also that vortex line density $\mathcal{L}$ is not too high,
\ this allows the relatively large loop to be radiated. They move slowly,
the smaller loops run down larger loop, then collide and reconnect with
them. So, outside of initial domain the well developed tangle is formed.
This, secondary, vortex tangle smoothly joins the initial tangle inside the
domain. This implies that in this case no boundary conditions are required
at all, and evolution of the vortex line density obeys equation (\ref%
{diffusion equation}) in infinite space with initial distribution $\mathcal{L%
}(\mathbf{r},0)$ inside this domain.

{\large 2. Radiation of loops}. The second situation can be realized when
the radiated vortex loops run away does not influence the initial vortex
tangle. It can happen, for instance if the initial tangle is very dense, so
it can radiate only very small loops, which propagate rapidly. These loops
run away without interaction with each other and with initial tangle where
they are radiated from. Other hypothetic variant appears when there is some
trap on the boundary absorbing vortex loops. Thus the vortex loops escaped
from the initial domain do not influence on the original vortex tangle. In
both case the boundary conditions can be found assuming that diffusive like
flux of length near boundary $\mathbf{J}=-D_{v}\nabla \mathcal{L}(x_{b},t)$
coincides with the flux executed by vortex loops radiated through the
(right) boundary $\mathbf{J}_{r}(x_{b},t)$. The latter is evaluated by small
modernization of (\ref{net flux})

\begin{equation}
\mathbf{J}_{r}=\int ln(l,x_{b}+R\cos \theta )(\mathbf{v}_{l}\cos \theta )P(R)%
\frac{\sin \theta d\theta d\varphi }{4\pi }dldR.  \label{flux rad}
\end{equation}

Contribution due to remote loops (term $R\cos \theta $ in argument for $%
n(l,x_{b}+R\cos \theta )$) is evaluated as early giving obvious result $%
\mathbf{J}=-(1/2)D_{v}\nabla \mathcal{L}(x_{b},t)$. To evaluate contribution
from the first one $\mathbf{J}_{evap}$\textbf{\ }we use again solution $%
n(l)=Cl^{-5/2}$ and relation (\ref{probability(x)}). Simple calculation lead
to result that flux $\mathbf{J}_{evap}$ of length due to evaporation
(radiation) is
\begin{equation}
\mathbf{J}_{evap}=C_{rad}\beta \mathcal{L}^{\frac{3}{2}},  \label{J evap}
\end{equation}%
with $C_{rad}\approx 0.47.$ The sum of $\mathbf{J}_{x}\,$\ and $\mathbf{J}%
_{evap}$ gives the flux through the boundary, which has to be equal to $%
-D_{v}\nabla \mathcal{L}(x_{b},t)$ at the boundary. Then we finally have the
following boundary condition
\begin{equation}
C_{rad}\mathcal{L}^{\frac{3}{2}}+\frac{1}{2}D_{v}\nabla \mathcal{L}%
(x_{b},t)=0.  \label{NBC}
\end{equation}

{\large 3. Solid walls.} In case of solid wall which corresponds to some
experiments the situation is more involved. Vortices can annihilate on the
solid wall, they can undergo pinning and depinning radiating vortices back
to the bulk of helium. Surely this requires a special treatment which goes
beyond the scope of the work. One the possible ways is to consider the solid
wall as a "partial" trap, which catches the loops and re-emits a part of
them back into the volume. Formally it can be written as condition (\ref{NBC}%
) with additional term $\mathbf{J}_{back}$ describing the back flux. Without
detailed analysis it can be supposed that the back flux is proportional to
the vortex line density on boundary $\mathbf{J}_{back}=-C_{back}\mathcal{L}%
(x_{b},t)$ with coefficient $C_{back}$ depending on dynamics of line on the
wall (jumps between pinning sites, Kelvin waves dynamics near the wall
etc.). Thus the boundary condition can be written in form.%
\begin{equation}
C_{rad}\mathcal{L}^{\frac{3}{2}}+\frac{1}{2}D_{v}\nabla \mathcal{L}%
(x_{b},t)-C_{back}\mathcal{L}=0.  \label{re-emission}
\end{equation}

\section{Decay of the vortex tangle}

Let us discuss the question how the approach developed can be applied to the
problem of decay of the vortex tangle at zero temperature. As it was
discussed in the Introduction, the contribution of diffusion (or radiation
of loops) is usually ignored, mainly due to smallness of the diffusion
constant offered in paper by Tsubota et all \cite{Tsub-diff}. Let us examine
this work more thoroughly. In paper \cite{Tsub-diff}~there was studied
one-dimensional evolution (spacial spreading) of the vortex tangle
concentrated initially in some domain of space and having nonuniform
distribution there as it is depicted in the first picture in Fig. 3.

\begin{figure}[tbp]
\includegraphics[width=7cm]{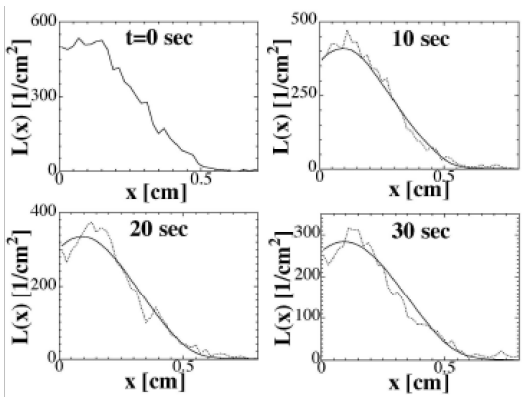}
\caption{Development of the
vortex line density distribution. The dotted line shows the results of the
numerical simulation of the diffusion equation with the diffusion constant is equal to $C_{d}\approx
0.1\ast 10^{-3}\ cm^{2}/s$ and with theauxiliary term $-\protect%
\chi _{2}(\protect\kappa /2\protect\pi )\mathcal{L}^{2}$, (Tsubota et, al.
\protect\cite{Tsub-diff}}
\label{}
\end{figure}

In the rest pictures distribution of the VLD at different moments of time is
shown.\ To describe this evolution of the VLD it had \ been supposed that
quantity $\mathcal{L}(x,t)$ obeys equation (\ref{diffusion equation}) with
additional term $-\chi _{2}(\kappa /2\pi )\mathcal{L}^{2}$ in the right hand
side. In turn this term was introduced to describe the decay of the vortex
tangle in previous numerical simulation made by Tsubota et al. \cite%
{Tsubota2000}. \emph{Because of this additional term contribution of
diffusion to the whole decay would be significantly underestimated. }We
would like to note that there is possible to choose another reasoning,
namely, to consider that decay of the vortex tangle in both cited works
occurs mainly by the diffusion process (with the diffusion coefficient
calculated in the present work). We calculated spatial -temporal evolution
of vortex tangle (under condition of numerical experiment by Tsubota et al.)
with the use of equation (\ref{diffusion equation}) Results depicted in Fig.
4. \ enables us to conclude that the approach developed describes
satisfactory the evolution of vortex tangle without any additional
supposition.

\begin{figure}[tbp]
\includegraphics[width=8cm]{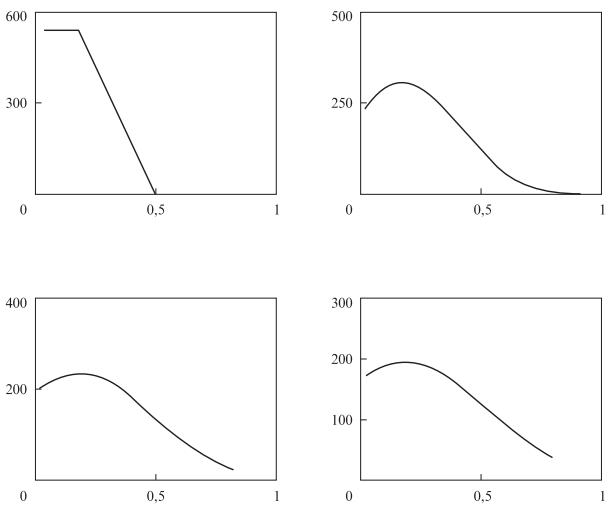}
\caption{Evolution of the vortex line
density calculated with use of equation (\protect\ref{diffusion equation})
without ani auxiliary term, the diffusion constant is equal to $C_{d}\approx
2.2\ast 10^{-3}\ cm^{2}/s$.}
\label{}
\end{figure}

Thus, we demonstrated that evolution of the vortex line density in numerical
experiment \cite{Tsubota2000} can be described in terms of pure diffusion of
the vortex tangle with the diffusion coefficient $C_{d}\approx 2.2\ast
10^{-3}\ cm^{2}/s$.

Let us now discuss two recent experiments on decay of the voretx tangle at
very small temperature \cite{Pickett-2006} and \cite{Golov07}. \ The authors
reported attenuation of vortex line density in superfluid turbulent helium, (%
$^{3}$He-B in paper \cite{Pickett-2006} and $^{4}$He in paper \cite{Golov07}%
). \ They attribute the decay of the vortex tangle to the classical
turbulence mechanisms. \ Without discussion of this variant we would like
just to estimate the contribution into attenuation of the vortex line
density due to the pure diffusion mechanism.

In the upper picture of Fig. 5 we displayed the Fig 2. of work \cite%
{Pickett-2006} showing results of measurements on the temporal behavior of
the average vortex line density $\mathcal{L}(t)$ (solid curves, see for
details \cite{Pickett-2006}). In the lower picture of Fig. 5 we depicted the
evolution the same quantity (for initial condition $\mathcal{L}=10^{8}\
1/cm^{2}$) due to diffusion process described in the present paper. The
initial domain of high vortex line density was created in the volume $^{3}$%
He-B, so its diffusion behavior should satisfy the first type boundary
condition. The straight line in the lower picture exactly corresponds to
line A in the \ upper one (which was named by authors of \cite{Pickett-2006}
as "limiting behavior"). It is easy to see that there is perfect agreement
between experimental data and theoretical predictions.

\begin{figure}[tbp]
\includegraphics[width=5cm]{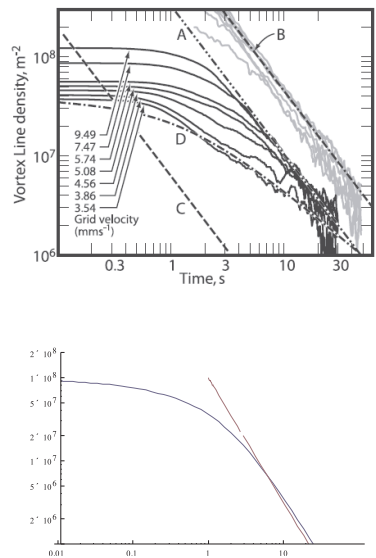}
\caption{Comparison of experimental data on the decay of
superfluid turbulence obtained in \protect\cite{Pickett-2006} (upper figure)
and our theoretical result (lower). We calculated temporal evolution of the
averaged vortex line density (for initial condition $\mathcal{L}=10^{8}\
1/cm^{2}$) due to diffusion process described in the present paper. The
initial domain of high vortex line density was created in the volume $^{3}$%
He-B, so its diffusion like behavior should satisfy the first type boundary
condition. The straight line in the lower figure exactly corresponds to line
A in the upper figure 5 (which was named by authors of \protect\cite%
{Pickett-2006} as "limiting behavior").}
\label{fig6}
\end{figure}

In contrast to work \cite{Pickett-2006} in the paper \cite{Golov07} the
decay of vortex tangle in He-II was observed in the closed cube with solid
walls. In the upper picture of Fig. 6 there is depicted the temporal
behavior of the average vortex line density$\mathcal{L}_{av}(t)$.

 We calculated the same
dependence on the base of diffusion equation (\ref{diffusion equation}),
result is shown in the lower picture of Fig. 6. \ As discussed above we can
( in the frame of the present paper) determine the boundary condition only
up to coefficient of re-emission $C_{back}$. Considering it as a fitting
parameter we have chosen the value of $C_{back}\approx $ $0.9$. It can be
seen that the decay of the vortex tangle due to diffusion reproduces some
feature observed in experiment.

\begin{figure}[tbp]
\includegraphics[width=5cm]{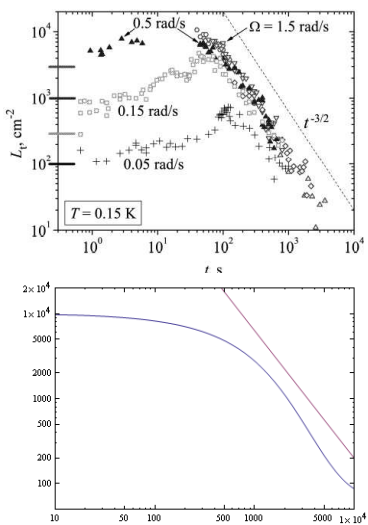}
\caption{Comparison of experimental data on the
decay of superfluid turbulence obtained in \protect\cite{Golov07} (upper
figure) and our theoretical result (lower). We calculated temporal evolution
of the averaged vortex line density (for initial condition $\mathcal{L}%
=10^{4}\ 1/cm^{2}$) due to diffusion process described in the present paper.
The superfluid turbulence in \protect\cite{Golov07} was created in the cubic
container with solid boundaries. Therefore we had chosen the third type
boundary condition with the fitting parameter $C_{back}\approx $ $0.9$. }
\label{}
\end{figure}

In particular, there is some plateau anticipatory decay of the tangle. Full
decay of the tangle occurs in time of the order of few thousands seconds as
it was observed in experiments. The slope of curve in interval of \ most
intensive decrease shows the dependence close to $\ \sim t^{-3/2}$ (The
straight line, its position coinsides with position of the straight line in
upper picture). \ Thus, there is again very good agreement between
experimental data and theoretical predictions.There is one very interesting
by-product of the consideration exposed above. It is easy to notice that for
small values of vortex line density the bend appears on curve $\mathcal{L}(t)
$. It corresponds to the fact that the diffusive-like flux of the length
vanishes (because of vanishing the gradient $\nabla \mathcal{L}(x_{b},t)$)$%
\mathcal{\ }$\ and the only first and third terms in the boundary condition (%
\ref{re-emission}) survive. Let us recall that these terms correspond to
radiation of loops from the bulk to the boundary and to the re-emission of
loops from the wall into bulk of helium. Comparing this terms (for the
chosen value $C_{back}\approx $ $0.9$) we obtain that equilibrium is reached
for value of the vortex line density $\mathcal{L}$ of order $50\div 100$\ $%
1/cm^{2}$. This value can be considered as a "background" value of
pre-existing vortices in helium

\section{Conclusion}

In summary, the theory describing evolution of inhomogeneous vortex tangle
at zero temperature was developed on the bases of kinetics of merging and
splitting vortex loops. Using the Gaussian model for vortex loops we
calculated the (size dependent) free path and mean quadratic velocity of
vortex loops. With the use of these quantities we calculated the flux of the
vortex line density $\mathcal{L}(x,t)$ in inhomogeneous vortex tangle and
demonstrated that under certain circumstances it satisfies to the diffusion
like equation with the coefficient equal approximately to $2.2\kappa $. We
used this equation to describe the decay of the vortex tangle at very low
temperature. We compare solution with the recent experiments on decay of the
superfluid turbulence. The good agreement with the experimental data allowed
us to conclude that the diffusion processes give the significant
contribution in the free decay of the vortex tangle at absence of normal
component.

\begin{center}
\textbf{ACKNOWLEDGMENTS}
\end{center}

This work was partially supported by grant 07-02-01124 from the RFBR and
grant of President Federation on the state support of leading scientific
schools RF NSH-6749.2006.8. \ I am grateful to participants of the LT
25(Amsterdam, 2008), especially Prof. M. Tsubota for useful discussion of
this work.

\end{document}